\documentclass[fleqn,usenatbib]{mnras}

\usepackage{times}

\usepackage[T1]{fontenc}

\DeclareRobustCommand{\VAN}[3]{#2}
\let\VANthebibliography\thebibliography
\def\thebibliography{\DeclareRobustCommand{\VAN}[3]{##3}\VANthebibliography}

\usepackage{graphicx}	
\usepackage{amsmath}	
\usepackage{amssymb}	

\newcommand{\be}{\begin{eqnarray}}
\newcommand{\ee}{\end{eqnarray}}

\title[The BH spins of MAXI~J1535 and 4U~1630]{The spins of the Galactic black holes in MAXI~J1535--571 and 4U~1630--472 from \textsl{Insight-HXMT}}

\author[Q. Liu et al.]{
Qichun Liu,$^{1}$
Honghui Liu,$^{1}$
Cosimo Bambi,$^{1}$\thanks{E-mail: bambi@fudan.edu.cn} 
and Long Ji$^{2}$
\\
$^{1}$Center for Field Theory and Particle Physics and Department of Physics, Fudan University, 200438 Shanghai, China\\
$^{2}$School of Physics and Astronomy, Sun Yat-sen University, 519082 Zhuhai, Guangdong, China
}

\date{\today}

\pubyear{2021}

\begin{document}
\label{firstpage}
\pagerange{\pageref{firstpage}--\pageref{lastpage}}
\maketitle

\begin{abstract}
\textsl{Insight-HXMT} is the first Chinese X-ray astronomical mission. With a 1-250~keV bandpass, an energy resolution of 150~eV at 6~keV, and without being subject to pile-up distortions, it is suitable to study bright X-ray sources like Galactic black holes. In the present paper, we study \textsl{Insight-HXMT} observations of the X-ray binaries MAXI~J1535--571 and 4U~1630--472 during their outbursts in 2017 and 2020, respectively. From the analysis of the reflection features, we are able to infer the black hole spin parameter in the two sources. For MAXI~J1535--571, we find a spin parameter very close to 1, $a_* = 0.9916 \pm 0.0012$ (90\% C.L., statistical error), which is consistent with the previous \textsl{NuSTAR} measurement. For 4U~1630--472, we find a moderately high value of the black hole spin parameter, $a_* =0.817 \pm 0.014$ (90\% C.L., statistical error), which is lower than the almost extremal value found with \textsl{NuSTAR} data.
\end{abstract}

\begin{keywords}
accretion, accretion discs -- black hole physics
\end{keywords}



\section{Introduction}

In 4-dimensional general relativity, black holes are relatively simple objects and completely characterized by only three parameters: the mass $M$, the spin angular momentum $J$, and the electric charge $Q$. This is the result of the no-hair theorem, which is valid under quite general and physically reasonable assumptions~\citep{1971PhRvL..26..331C,1975PhRvL..34..905R,2012LRR....15....7C}. Astrophysical black holes are normally expected to have a negligible electric charge~\citep{2017bhlt.book.....B}, and therefore they should be described only by the mass and the spin angular momentum. The mass of a black hole can be inferred by studying the motion of material orbiting the compact object; in the case of black holes in X-ray binaries, we can study the orbital motion of the companion star \citep{2014SSRv..183..223C}. Measurements of black hole spin are more challenging: the spin has no gravitational effects in Newtonian gravity and therefore spin measurements require the analysis of relativistic effects occurring in the strong gravity region close to the black hole event horizon. On the other hand, accurate black hole spin measurements are important to understand the formation and the evolution of these systems \citep{Reynolds2014,2019NatAs...3...41R,2021SSRv..217...65B}. In the case of stellar-mass black holes, the spin is generally thought to reflect the formation mechanism of the object. For supermassive black holes in galactic nuclei, the black hole spin is thought to be determined by the merger history of the host galaxy as well as by the accretion history of the compact object.

X-ray reflection spectroscopy is currently the most popular method for measuring the spins of accreting black holes~\citep{Reynolds2014,Brenneman2013,2021SSRv..217...65B}. So far it has provided the spin measurement of over twenty stellar-mass black holes and about forty supermassive black holes~\citep{2021SSRv..217...65B}. The method relies on the analysis of the reflection spectrum generated from a cold and thin disk illuminated by a hot corona. The thermal spectrum of geometrically thin and optically thick accretion disks is peaked in the soft X-ray band for stellar-mass black holes and in the UV band for the supermassive ones. Thermal photons from the inner part of the accretion disk can inverse Compton scatter off free electrons in the corona, which is some hotter ($\sim$100~keV) plasma close to the black hole. A fraction of the Comptonized photons can illuminate the disk: Compton scattering and absorption followed by fluorescent emission generate the reflection spectrum~\citep{Ross2005,Garc2010}. One of the most prominent features in the reflection spectrum is normally the iron K$\alpha$ complex, which is around 6.4~keV in the case of neutral or weakly ionized iron atoms and can shift up to 6.97~keV in the case of H-like iron ions. Since the iron K$\alpha$ complex is an intrinsically narrow feature, it is particularly suitable to measure the effects of relativistic blurring, estimate black hole spins, and even test fundamental physics \citep[see, e.g.,][]{2017RvMP...89b5001B,Tripathi2021a}.

\textsl{Insight-HXMT} is the first Chinese X-ray astronomical mission \citep{Zhang2014}. It was successfully launched in June 2017. It has three slat-collimated instruments: the Low Energy X-ray Telescope (LE, 1-15~keV), the Medium Energy X-ray Telescope (ME, 5-30~keV), and the  High Energy X-ray Telescope (HE, 20-250~keV) \citep[][]{Huang2018}. The energy resolution of LE is 150~eV at 6~keV. The three instruments are hardly subject to pile-up distortions, which makes them suitable to observe bright sources. In this work, we present the analysis of \textsl{Insight-HXMT} observations of the Galactic black holes MAXI~J1535--571 and 4U~1630--472. We select reflection dominated spectra of these sources and we are able to measure the spin of the two black holes. The measurements of the black hole spins of MAXI~J1535--571 and 4U~1630--472 were previously reported by \citet{Xu2018} and \citet{King2014}, respectively, in both cases by analyzing \textsl{NuSTAR} data \citep[][]{Harrison2013}. Our work confirms the very high spin of the black hole in MAXI~J1535--571 found by \citet{Xu2018}, while we do not find a very high spin parameter for the black hole in 4U~1630--472 as reported in \citet{King2014}.

The paper is organized as follows. In Section~\ref{ob}, we present the sources and the observations of our study. In Section~\ref{fit}, we report our spectral analysis. Discussion and conclusions are in Section~\ref{ed}.

\section{Observations and Data Reduction}
\label{ob}

Tab.~\ref{HXMT_BHobs} shows the list of \textsl{Insight-HXMT} observations of black hole binaries as of 30 September 2021. Our goal is to select all possible \textsl{Insight-HXMT} data suitable to measure black hole spins from the analysis of reflection features and without an existing spin measurement in the literature. Unfortunately, most observations are unsuitable for this objective.

There are several observations of Cyg~X-1 and GRS~1915+105, but there are no spectra with strong reflection features. In the case of MAXI~J1348--630 and MAXI~J1820+070, there are some \textsl{Insight-HXMT} spectra with reflection features, but the iron line is weak and/or not very broad, and eventually it is not possible to constrain the black hole spin. For GX~339--4, there are several \textsl{Insight-HXMT} observations of the 2021 outburst of this source, and their spectral analysis will be reported in another study (Liu et al., in preparation). Among the 42 observations of EXO~1846--031, there is only one observation showing reflection features, but it is too short and it is impossible to constrain the spin of the black hole. The spectra of Swift~J1658.2--4242 are contaminated by some nearby sources, which cannot be removed and prevent any accurate analysis of the reflection features of this source. In the end, we found that black hole spin measurements from the analysis of relativistically blurred reflection features are only possible for MAXI~J1535--571 and 4U~1630--472. In particular, we found 8~observations of MAXI~J1535--571 and 5~observations of 4U~1630--472 suitable for our study.

Tab.~\ref{HXMT_spins} lists the black hole spin measurements obtained by the analysis of \textsl{Insight-HXMT} observations to date. There are two spin measurements obtained from the continuum-fitting method, namely the analysis of the thermal spectrum of the disk~\citep{1997ApJ...482L.155Z,2014SSRv..183..295M}. For GRS~1915+105, an estimate of the black hole spin has been inferred using QPOs, but there is not yet a common consensus on the exact mechanism responsible for QPOs, so that measurement should be taken with caution even if it agrees with previous measurements obtained from the continuum-fitting and the iron line methods. For MAXI~J1535--571, \citet{2020JHEAp..25...29K} analyze only one of the 8 \textsl{Insight-HXMT} observations of MAXI~J1535--571, so they have a very low statistics.

\begin{table*}
	\centering
	\caption{Summary of the \textsl{Insight-HXMT} observations of black hole binaries as of 30 September 2021.}
	\label{HXMT_BHobs}
	\begin{tabular}{lrr} 
		\hline\hline
		Source & Number of observations & Total exposure time~(ks)\\
		\hline
		4U~1543--47 & 51 & 1070 \\
		4U~1630--472 & 55 & 700\\
		Cyg~X-1 &  140 & 1840  \\
		EXO~1846--031 & 42 & 800 \\
		GRS~1716--249 & 2 & 250 \\
		GRS~1915+105 & 150 & 2600 \\
		GX~339--4 & 128 & 1930 \\
		H~1743--322 & 40 & 440 \\
		MAXI~J1348--630 & 126 & 2620 \\
		MAXI~J1535--571 & 18 & 430 \\
		MAXI~J1543--564 & 1 & 80 \\
		MAXI~J1727--203 & 3 & 30 \\
		MAXI~J1820+070 & 146 & 2560 \\
		Swift~J1658.2--4242 & 23 & 470 \\
		Swift~J1728.9--3613 & 23 & 250 \\
		\hline\hline
	\end{tabular}\\
\end{table*}

\begin{table*}
	\centering
	\caption{Summary of black hole spin measurements with \textsl{Insight-HXMT} data as of 30 September 2021. CFM = continuum-fitting method; XRS = X-ray reflection spectroscopy; QPOs = Quasi-periodic oscillations.}
	\label{HXMT_spins}
	\begin{tabular}{lccc} 
		\hline\hline
		Source & Spin measurement & Method & Reference \\
		\hline
		Cyg~X-1 & $> 0.967$ (3-$\sigma$) & CFM & \citet{2020JHEAp..27...53Z}  \\
		GRS~1915+105 & $0.99836 \pm 0.00028$ (90\% C.L.) & QPOs & \citet{2021ApJ...909...63L} \\
		GX~339--4 & $> 0.88$ (90\% C.L.) & XRS & Liu et al., in preparation \\
		MAXI~J1535--571 & $0.7 _{-0.3}^{+0.2}$ (90\% C.L.) & XRS & \citet{2020JHEAp..25...29K} \\
		MAXI~J1820+070 & $0.2_{-0.3}^{+0.2}$ (1-$\sigma$) & CFM & \citet{2021MNRAS.504.2168G} \\
		MAXI~J1820+070 & $0.14 \pm 0.09$ (1-$\sigma$) & CFM & \citet{2021ApJ...916..108Z} \\
		\hline\hline
	\end{tabular}\\
\end{table*}

\subsection{Observations}

The X-ray binary transient MAXI J1535--571 was discovered by \textsl{MAXI} \citep[][]{Matsuoka2009} on 2 September 2017 \citep[][]{Nakahira2018}. During the outburst, extremely bright radio and sub-mm counterparts were detected, suggesting that the source is a low-mass X-ray binary with a black hole \citep{Sridhar2019}. From the analysis of \textsl{AstroSat} data, \citet{Sridhar2019} inferred the black hole mass $M = \left(10.4 \pm 0.6 \right)~M_\odot$ and the distance from us $D = 5.4_{-1.1}^{+1.8}$~kpc.

4U~1630--472 is a recurrent black hole X-ray binary. It was discovered by the \textsl{Vela 5B} and \textsl{Uhuru} satellites in 1969 \citep{1986Ap&SS.126...89P,1976ApJ...210L...9J} and has regular outbursts of 100-200~days followed by quiescent periods of about 500~days \citep{Tetarenko:2015vrn}. The hydrogen column density along its line of sight is very high, probably close to 10$^{23}$~cm$^{-2}$ \citep{Gatuzz:2018nog}, which prevents a dynamical measurement of the mass of the black hole from the study of the motion of the companion star. From the scaling of the photon index with the mass accretion rate, \citet{2014ApJ...789...57S} inferred a black hole mass around 10~$M_\odot$. IR observations suggest a distance of 10-11~kpc from us \citep{2001A&A...375..447A,2014ApJ...789...57S}.

\begin{figure}
	\includegraphics[width=0.97\columnwidth]{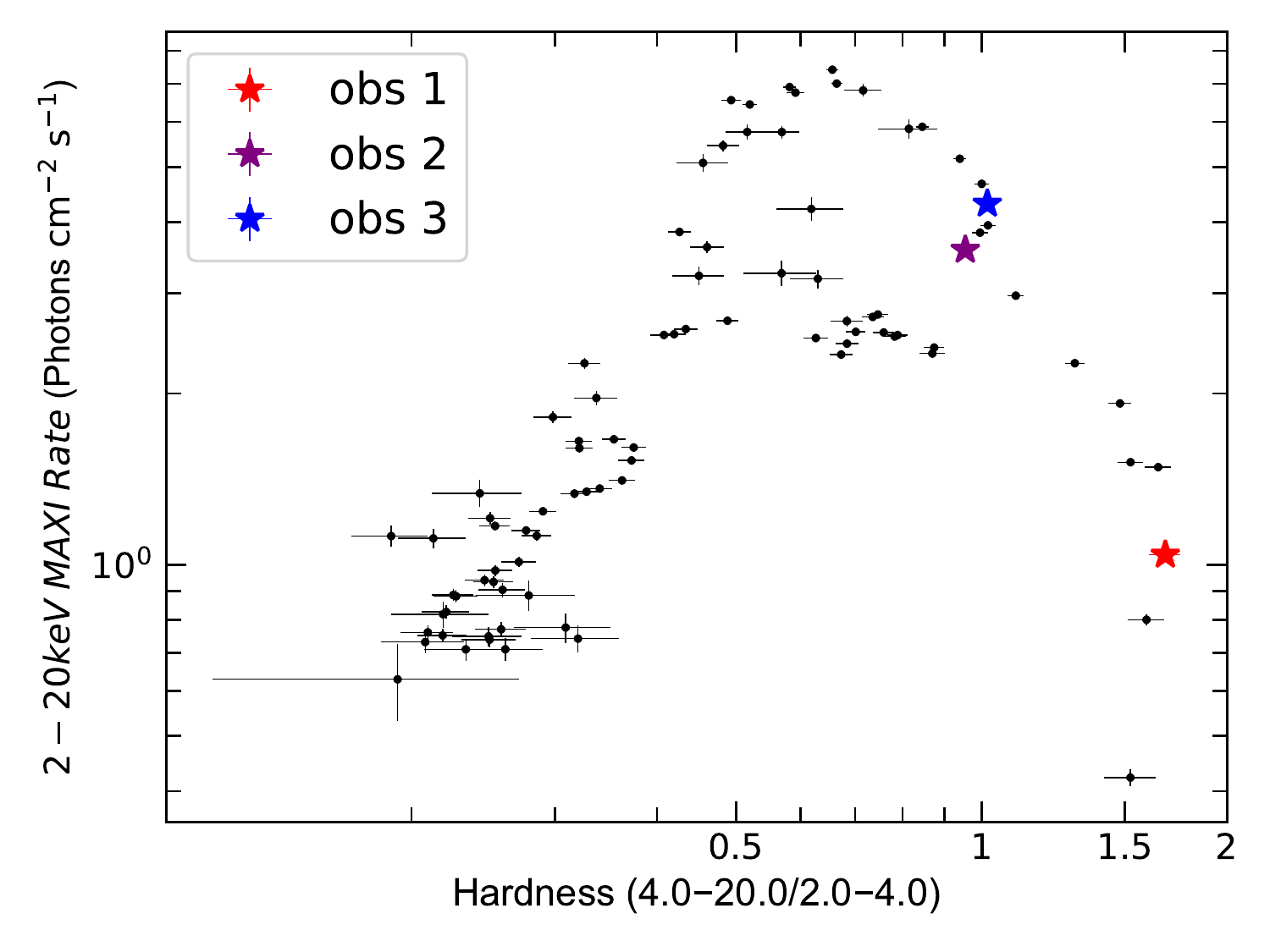}
	\vspace{-0.2cm}
    \caption{HID for the 2017 outburst of MAXI~J1535--571 from \textsl{MAXI}/GSC (2-20~keV). The red, purple, and blue stars mark the observations analyzed in this work.}
    \label{MAXI_HID}
\vspace{0.5cm}
	\includegraphics[width=0.97\columnwidth]{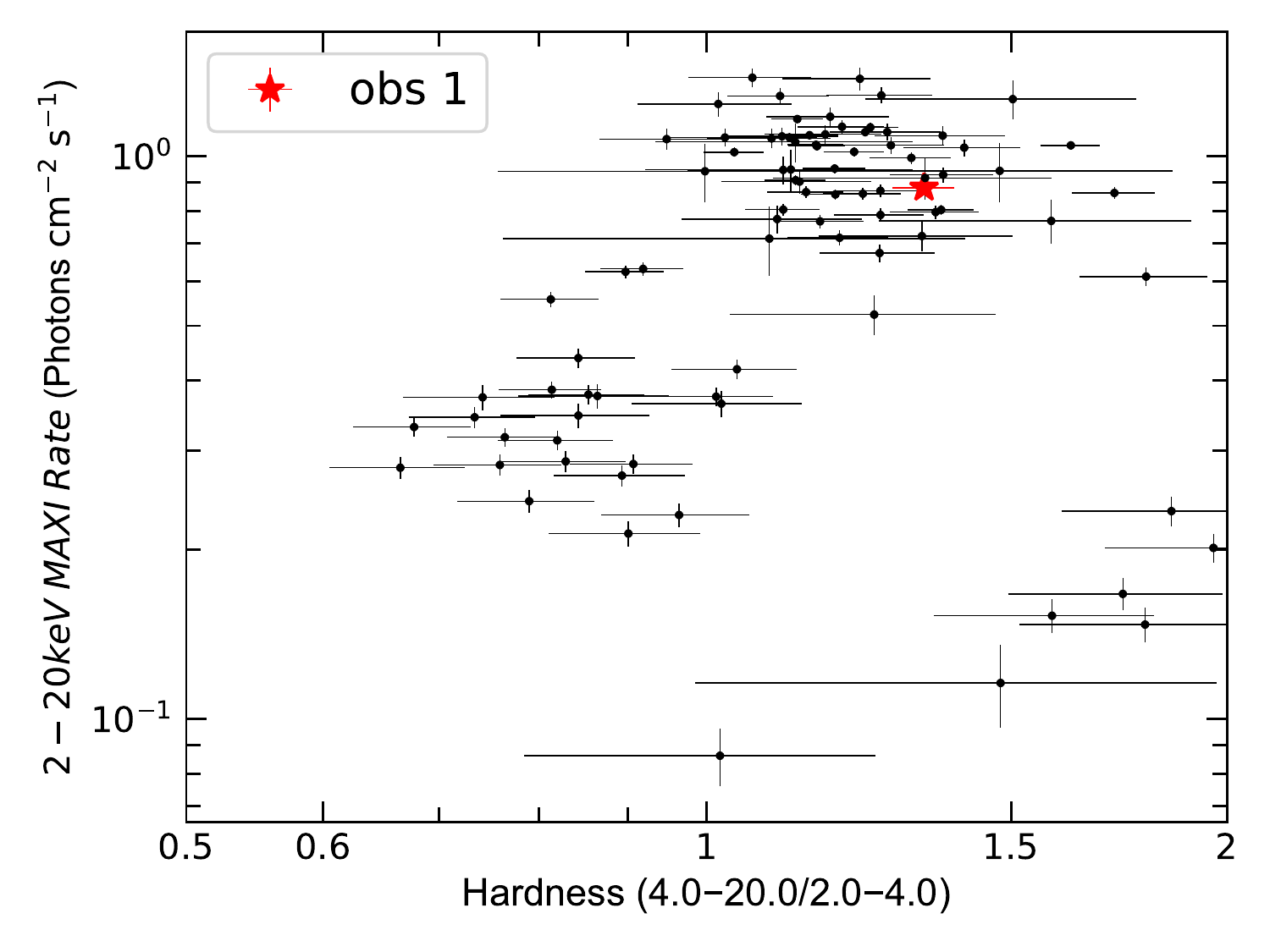}
	\vspace{-0.2cm}
    \caption{HID for the 2020 outburst of 4U~1630--472 from \textsl{MAXI}/GSC (2-20~keV). The red star marks the observation analyzed in this work.}
    \label{4U_HID}
\end{figure}

Tab.~\ref{observation_table} shows the observations of MAXI~J1535--571 and 4U~1630--472 analyzed in our work. Our selection criterion is based on the equivalent width of the iron line, which we require to be larger than 50~eV. For MAXI~J1535--571, we found 8~\textsl{Insight-HXMT} observations (out of 18) with strong reflection features that meet our selection criterion. However, some observations are consecutive or almost consecutive. After checking the absence of variability in the flux and hardness of the source in the consecutive observations (see the fourth and fifth columns in Tab.~\ref{observation_table}), we merged them together into three spectra as shown in Tab.~\ref{observation_table}. Fig.~\ref{MAXI_HID} shows the hardness-intensity diagram (HID) of the 2017 outburst of the source obtained from the \textsl{MAXI} observations and the location of our three \textsl{Insight-HXMT} spectra. Obs~1 is at the end of hard state and overlaps with the \textsl{NuSTAR} observation analyzed in \citet{Xu2018}. In obs~2 and obs~3, the source is in the intermediate state. In the case of 4U~1630--472, we found 5 observations (out of 51) with some reflection features. These observations are short and almost consecutive. Since the source does not show any variability in flux and hardness during the two day period of these observations (see, again, the fourth and fifth columns in Tab.~\ref{observation_table}), we merged them into a single spectrum. Fig.~\ref{4U_HID} shows the HID of the 2020 outburst of 4U~1630--472 as inferred from \textsl{MAXI} and the red start shows the position of the \textsl{Insight-HXMT} spectrum. The source was in the intermediate state.

\begin{table*}
	\centering
	\caption{Summary of the \textsl{Insight-HXMT} observations of MAXI~J1535--571 and 4U~1630--472 analyzed in the present work. The observation IDs are in the form P011453500XXX (for MAXI~J1535--572) or P020503000XXX (for 4U~1630--472). In the table, we only report the last three digits of the observation IDs. The last two columns report, respectively, the exposure and the LE count rate of the merged spectra.}
	\label{observation_table}
	\begin{tabular}{lclllllc} 
		\hline\hline
		Source & Obs & Obs ID & Count rate (s$^{-1}$) & Hardness & Obs date & LE/ME/HE exposure (ks) & LE count rate (s$^{-1}$) \\
		& & & (2-10~keV) & (4-10~keV/2-4~keV) & & & \\
		\hline
		MAXI~J1535--571 & obs~1 & 101 & 243.2 & 1.16& 2017 Sept 6 & 4.61/6.84/3.38 & 255.9 \\
		 & & 102 & 251.1 &1.15& 2017 Sept 6&\\
		 & &103 &262.6 &1.15& 2017 Sept 6&\\
		 & & 106 & 294.5 &1.14& 2017 Sept 6\\
		 & & 107 & 301.4 &1.15& 2017 Sept 7\\
		 & obs~2 & 144 & 1002 &0.78 & 2017 Sept 12 & 4.32/10.37/11.01 & 981.7 \\
		 & & 145 & 1020 & 0.78&2017 Sept 12&\\ 
		 & obs~3 &  301 &1127& 0.83 & 2017 Sept 15 & 0.60/3.38/1.64 & 1085 \\
		\hline
		4U~1630--472 & obs~1 & 101& 55.9& 1.17& 2020 Mar 19 & 13.87/17.95/17.23 & 56.03 \\
		& & 102 & 56.6 & 1.18&2020 Mar 19\\
		& & 201 & 57.8 & 1.17&2020 Mar 20\\
		& & 202 & 58.9& 1.18&2020 Mar 20\\
		& & 203 &59.2 & 1.19&2020 Mar 20\\ 
		\hline\hline
	\end{tabular}
\end{table*}

\subsection{Data Reduction}\label{sec:maths}
 
Following the official user guide\footnote{\href{http://hxmten.ihep.ac.cn/SoftDoc.jhtml}{http://hxmten.ihep.ac.cn/SoftDoc.jhtml}}, we extracted spectra using the software HXMTDAS v2.04 and the pipeline\footnote{\href{http://code.ihep.ac.cn/jldirac/insight-hxmt-code-collection/-/blob/master/version2.04/HXMT_analysis_2.04_2020.py}{http://code.ihep.ac.cn/jldirac/insight-hxmt-code-collection/-/blob/master/version2.04/HXMT\_analysis\_2.04\_2020.py}}. We also used the latest calibration database CALDBV v2.05. The criteria for estimating good time intervals is as follows: 1) the elevation angle is larger than 10~degree; 2) the geomagnetic cutoff
rigidity is larger than 8~GeV; 3) the pointing offset angle is smaller than 0.1~degree; 4) at least 300~s away from the South Atlantic Anomaly (SAA). The spectral backgrounds were calculated by using the tools LEBKGMAP, MEBKGMAP, and HEBKGMAP \citep{Liao2020,Guo2020,2020JHEAp..27...14L}.

\section{Spectral analysis}
\label{fit}

For the spectral analysis of MAXI~J1535--571 and 4U~1630--472, we consider the energy bands 2-8~keV (LE), 10-20~keV (ME), and 30-80 keV~(HE). The XSPEC v12.10.1s software package \citep[][]{Arnaud1996} is used in our work. We bin the \textsl{Insight-HXMT} spectra in order to reach a minimal signal-to-noise ratio of 25. We adopt abundances from \citet{Wilms2000} and cross sections from \citet{Verner1996}. In what follows, the uncertainties on the estimates of the model parameters are at the 90\% confidence level and only include the statistical errors.

\subsection{MAXI~J1535--571}

\begin{figure}
	\includegraphics[width=0.97\columnwidth]{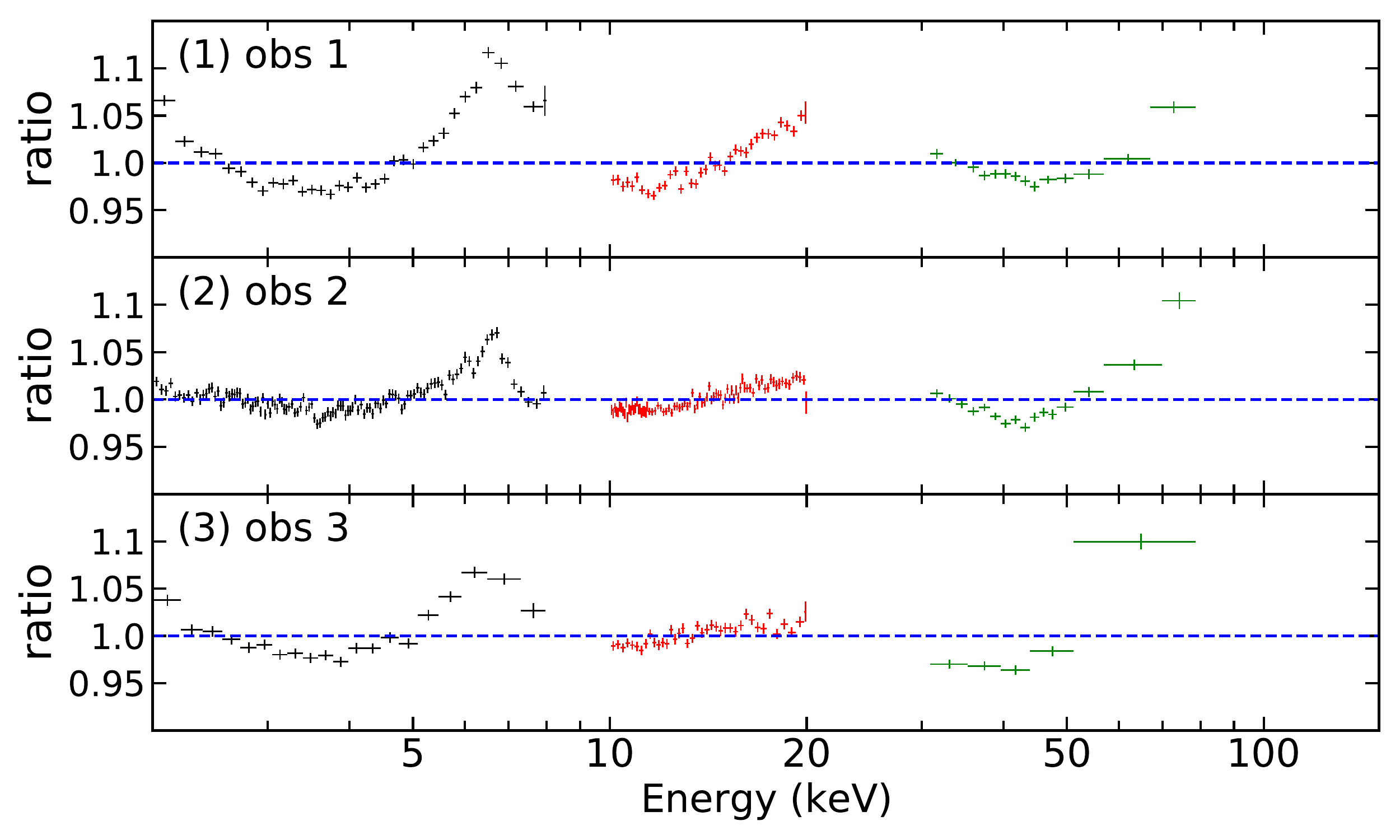}
	\vspace{-0.2cm}
    \caption{Data to best-fit model ratios for obs~1 (top quadrant), obs~2 (central quadrant), and obs~3 (bottom quadrant) of Model~0 of MAXI~J1535--571. Black, red, and green crosses are, respectively, for LE, ME, and HE data. In all observations we see a broad iron line peaked at 6-7~keV.}
    \label{fig:MAXI_iron_line_figure}
\end{figure}

First, we fit the three spectra together with an absorbed continuum from thermal Comptonization (Model~0); in XSPEC language, the model is \texttt{constant*tbabs*nthcomp}. The multiplicative constant is the cross-calibration constant among the three instruments, LE, ME, and HE. \texttt{tbabs} describes the Galactic absorption \citep[][]{Wilms2000} and has only one free parameter, the hydrogen column density, $N_{\rm H}$. \texttt{nthcomp} describes the Comptonized photons from the corona \citep[][]{Zdziarski1996,Zycki1999}. We have three free parameters in \texttt{nthcomp}: the temperature of the seed photons from the disk, $kT_{\rm bb}$, the temperature of the electron in the corona, $kT_{\rm e}$, and the power-law photon index, $\Gamma$. Fig.~\ref{fig:MAXI_iron_line_figure} shows the resulting data to best-fit ratio, where we can see a broad iron line peaked at 6-7~keV in the three spectra. The best-fit has $\chi_{\nu}^2=8245.95/2529=3.26056$.

To improve the fit, we add the model \texttt{diskbb} \citep[][]{Mitsuda1984}, which describes the thermal spectrum of a thin accretion disk. This is our Model~1. We link the value of the inner temperature of the disk $kT_{\rm in}$ in \texttt{diskbb} with the temperature of the seed photons $kT_{\rm bb}$ in {\tt nthcomp}. We find $\chi_{\nu}^2=6703.98/2526=2.65399$, and thus $\Delta\chi^2=-1541.97$ with respect to Model~0.

For our Model~2, we add the reflection spectrum of the accretion disk. We use \texttt{relxillCp}~\citep{Dauser2013,2013ApJ...768..146G,2014ApJ...782...76G} and our total model is \texttt{constant*tbabs*(diskbb+relxillCp)}. \texttt{relxillCp} includes \texttt{nthcomp} when the reflection fraction $R_{\rm f}$ is positive, so we remove \texttt{nthcomp} from the total model. The spacetime is described by the dimensionless black hole spin parameter $a_*$, while the black hole mass does not directly enter the calculations of the reflection spectrum and therefore it is not a parameter of \texttt{relxillCp}. The black hole spin $a_*$, the inclination angle of the disk with respect to our line of sight $i$, and the iron abundance of the disk $A_{\rm Fe}$ are left free in the fit and we impose they do not change values over the three spectra. The inner edge of the accretion disk $R_{\rm in}$ is assumed to be at the innermost stable circular orbit (ISCO) of the spacetime, so it is not a free parameter in our fit and depends only on the value of the black hole spin $a_*$, as it is measured in units of the black hole gravitational radius $r_{\rm g} =G_{\rm N}M/c^2$. The outer edge of the accretion disk $R_{\rm out}$ is fixed to 400~$r_{\rm g}$. For the emissivity profile of the accretion disk, we adopt a broken power-law, so we have three parameters: the inner emissivity index $q_{\rm in}$, the outer emissivity index $q_{\rm out}$, and the breaking radius $R_{\rm br}$. For obs~1, we find that a simple power-law is enough, so we set $q_{\rm out} = q_{\rm in}$ and we have only one free parameter. For obs~2 and obs~3, we find that a broken power-law provides a better fit than a power-law and we thus leave $q_{\rm in}$, $q_{\rm out}$, and $R_{\rm br}$ free. The ionization parameter of the disk $\xi$ is left free in the fit and allowed to vary over obs~1, 2, and 3. Since \texttt{relxillCp} includes \texttt{nthcomp}, the model has also the photon index $\Gamma$ and the temperature of the electrons in the corona $kT_{\rm e}$.

\begin{table*}
	\centering
	\renewcommand\arraystretch{1.2}
	\caption{Summary of the best-fit values of Model~2, Model~2d, and Model~3 of MAXI~J1535--571. The reported uncertainties correspond to the 90\% confidence level for one relevant parameter ($\Delta\chi^2 = 2.71$). We note that $q_{\rm in}$ is allowed to vary from 0 to 10 in the fit, $A_{\rm Fe}$ from 0.5 to 10, $\log n_{\rm e}$ from 15 to 19, and $kT_{\rm e}$ from 1 to 400~keV; in some observations, the fit is stuck at one of the boundaries of the parameter and there is no upper/lower uncertainty.}
	\label{MAXI_model_2}
	\resizebox{18cm}{!}{
	\begin{tabular}{lccccccccc}
		\hline\hline
		& \multicolumn{3}{c}{Model~2} & \multicolumn{3}{c}{Model~2d} & \multicolumn{3}{c}{Model~3} \\
		Parameter & {obs~1} & {obs~2} & {obs~3} & {obs~1} & {obs~2} & {obs~3} & {obs~1} & {obs~2} & {obs~3} \\
		\hline
		{\tt tbabs} && \\
		$N_{\rm H}$ [10$^{22}$ cm$^{-2}$] 
		& $6.39^{+0.21}_{-0.14}$ & $6.05^{+0.07}_{-0.05}$ & $7.19^{+0.24}_{-0.12}$ 
		& $7.32_{-0.11}^{+0.12}$ & $ 6.53_{-0.09}^{+0.08}$ & $7.47_{-0.12}^{+0.20}$
		& $ 6.40^{+0.19}_{-0.22}$ & $5.99^{+0.09}_{-0.07}$ & $7.12 ^{+0.26}_{-0.19}$ \\
		\hline
		{\tt diskbb} && \\
		$T_{\rm in}$ [keV] 
		& $0.269 ^{+0.016}_{-0.016}$ & $ 0.225^{+0.015}_{-0.014}$ & $0.230^{+0.014}_{-0.014}$ 
		& $0.204_{-0.011}^{+0.005}$ & $0.198_{-0.006}^{+0.008}$ & $0.193_{-0.010}^{+0.006}$
		& $ 0.269 ^{+0.014}_{-0.015}$ & $ 0.220^{+0.015}_{-0.014}$ & $0.227^{+0.014}_{-0.015}$ \\
		${\rm Norm}$ 
		& $8^{+5}_{-3}\cdot 10^5$ &$1.1^{+0.6}_{-0.5}\cdot 10^7$&$2.4^{+2.0}_{-0.9}\cdot 10^7$ 
		& $ 1.7_{-0.6}^{+0.9}\cdot 10^7$ & $5.0_{-1.7}^{+1.6}\cdot 10^7$ & $16_{-5}^{+10}\cdot 10^7$
		& $7.8^{+4.7}_{-2.5}\cdot 10^5$ & $1.3^{+1.2}_{-0.6}\cdot 10^7$ & $2.7^{+2.1}_{-1.1}\cdot10^7$\\
		\hline
		{\tt relxillCp} && &\\
		{\tt relxillD} && &\\
		$q_{\rm in}$ 
		& $1.83^{+0.07}_{-0.07}$ &$10.0_{-0.3} $ & $10.0_{-0.9}$ 
		& $1.99_{-0.04}^{+0.04}$ & $10.00_{-0.15}$ & $10.00_{-0.17}$
		& $1.79^{+0.07}_{-0.07}$ &$10.0_{-1.4} $ & $9.8_{-1.6}^{+0.2}$  \\
		$q_{\rm out}$ & $= q_{\rm in}$ & $0.8^{+0.3}_{-0.4} $ & $1.1^{+0.4}_{-1.2} $ 
		& $= q_{\rm in}$ & $1.12_{-0.14}^{+0.10}$ & $0.19_{-0.09}^{+0.16}$
		& $= q_{\rm in}$ & $0.6^{+0.5}_{-0.6} $ & $ 1.0^{+0.6}_{-1.0} $\\
		$R_{\rm br}$~$[r_{\rm g}]$ 
		& -- & $4.1_{-0.4}^{+1.2} $ & $3.6  _{-0.6}^{+1.6} $ 
		& -- & $3.65_{-0.06}^{+0.26}$ & $5.5_{-0.2}^{+0.3}$
		& -- & $3.8_{-1.5}^{+1.8} $ & $3.3_{-0.9}^{+3.1} $\\
		$a_*$ 
		& \multicolumn{3}{c}{$0.9916_{-0.0012}^{+0.0012}$} 
		& \multicolumn{3}{c}{$0.996 _{-0.002}^{+0.002}$}
		& \multicolumn{3}{c}{$0.9978_{-0.0081}^{+0.0002}$}  \\
		$i$ [deg] 
		& \multicolumn{3}{c}{$74.2^{+0.6}_{-0.7}$}
		& \multicolumn{3}{c}{$74.2_{-0.8}^{+0.7}$}
		& \multicolumn{3}{c}{$76.4^{+0.6}_{-2.1}$} \\
		$\alpha_{13}$ 
		& \multicolumn{3}{c}{--}
		& \multicolumn{3}{c}{--}
		& \multicolumn{3}{c}{$-0.24^{+0.12}_{-0.04}$}\\
		$A_{\rm Fe}$ 
		& \multicolumn{3}{c}{$0.51^{+0.05}_{-0.01}$}
		& \multicolumn{3}{c}{$0.50^{+0.02}$}
		& \multicolumn{3}{c}{$0.50^{+0.04}$} \\
		$\log\xi$ [erg~cm~s$^{-1}$] 
		& $3.02^{+0.05}_{-0.08}$ & $2.43^{+0.05}_{-0.04}$ & $2.69^{+0.03}_{-0.06}$ 
		& $2.13_{-0.03}^{+0.04}$ & $2.30_{-0.07}^{+0.02}$ & $2.40_{-0.09}^{+0.04}$
		& $3.02  ^{+0.05}_{-0.08}$ & $2.43^{+0.05}_{-0.04}$ & $2.70^{+0.02}_{-0.05}$  \\
		${\Gamma}$ 
		& $1.89^{+0.02}_{-0.02}$ & $2.571 _{-0.004}^{+0.004} $ & $2.581 ^{+0.006}_{-0.008}$
		& $2.152_{-0.006}^{+0.005}$ & $2.690_{-0.003}^{+0.005}$ & $2.672_{-0.008}^{+0.005}$
		& $1.89^{+0.02}_{-0.02}$ & $2.570 _{-0.008}^{+0.004} $ & $2.579 ^{+0.008}_{-0.010}$\\
		$kT_{\rm e}$ [keV]  
		& $36^{+3}_{-2}$ &  $400_{-132}$ &  $ 400_{-128}$ 
		& -- & -- & --
		& $36^{+3}_{-3}$ &  $400_{-64}$ &  $ 400_{-114}$\\
		$E_{\rm cut}$ [keV]
		& -- & -- & --
		& $300^*$ & $300^*$ & $300^*$
		& -- & -- & -- \\
		$\log n_{\rm e}$ [cm$^{-3}$]
		& $15^*$ & $15^*$ & $15^*$
		& $15.000^{+0.009}$ & $ 18.02_{-0.15}^{+0.49}$ & $18.08_{-0.91}^{+0.02}$
		& $15^*$ & $15^*$ & $15^*$ \\
		$R_{\rm f}$ 
		& $0.35^{+0.05}_{-0.04}$ &  $ 0.319  ^{+0.011}_{-0.012}$ &  $0.52^{+0.03}_{-0.03}$ 
		& $0.584_{-0.015}^{+0.015}$ & $0.52_{-0.10}^{+0.02}$ & $0.72_{-0.03}^{+0.04}$
		& $0.36^{+0.06}_{-0.05}$ &  $ 0.320^{+0.015}_{-0.017}$ &  $0.52^{+0.04}_{-0.05}$\\
		${\rm Norm}$ 
		& $0.053_{-0.006}^{+0.007} $ &$0.530 _{-0.005}^{+0.005} $&$0.559_{-0.017}^{+0.016} $
		& $0.0871_{-0.0007}^{+0.0007}$ & $0.767_{-0.007}^{+0.010}$ & $0.655_{-0.013}^{+0.012}$
		& $0.053_{-0.003}^{+0.005} $ &$0.528 _{-0.009}^{+0.008} $&$0.554_{-0.019}^{+0.016} $ \\
		\hline
		{\tt constant} &&\\
		${\rm LE}$ & $1^*$ & $1^*$ & $1^*$ & $1^*$ & $1^*$ & $1^*$ & $1^*$ & $1^*$ & $1^*$ \\
		${\rm ME}$ 
		& $1.043^{+0.013}_{-0.009}$ & $0.962^{+0.006}_{-0.004}$ & $0.981^{+0.011}_{-0.006}$ 
		& $1.027_{-0.006}^{+0.005}$ & $0.961_{-0.004}^{+0.002}$ & $0.954_{-0.007}^{+0.005}$
		&$1.040^{+0.014}_{-0.008}$ & $0.962^{+0.004}_{-0.004}$ & $0.980^{+0.009}_{-0.005}$\\
		${\rm HE}$
		& $1.215^{+0.020}_{-0.018}$ & $0.954^{+0.007}_{-0.008}$ & $1.013^{+0.013}_{-0.013}$ 
		& $1.180_{-0.013}^{+0.013}$ & $0.980_{-0.007}^{+0.004}$ & $1.008_{-0.012}^{+0.008}$
		& $1.208^{+0.020}_{-0.017}$ &$0.954^{+0.008}_{-0.008}$ &$1.013^{+0.015}_{-0.012}$  \\
		\hline
		$\chi^2/\nu$ 
		& \multicolumn{3}{c}{2667.43/2504} 
		& \multicolumn{3}{c}{2790.28/2504}
		& \multicolumn{3}{c}{2663.40/2503} \\
		&  \multicolumn{3}{c}{=1.06527} 
		& \multicolumn{3}{c}{=1.11433}
		& \multicolumn{3}{c}{=1.06408} \\
		\hline\hline
	\end{tabular}
	}
\end{table*}

\begin{figure*}
	\includegraphics[width=1.97\columnwidth]{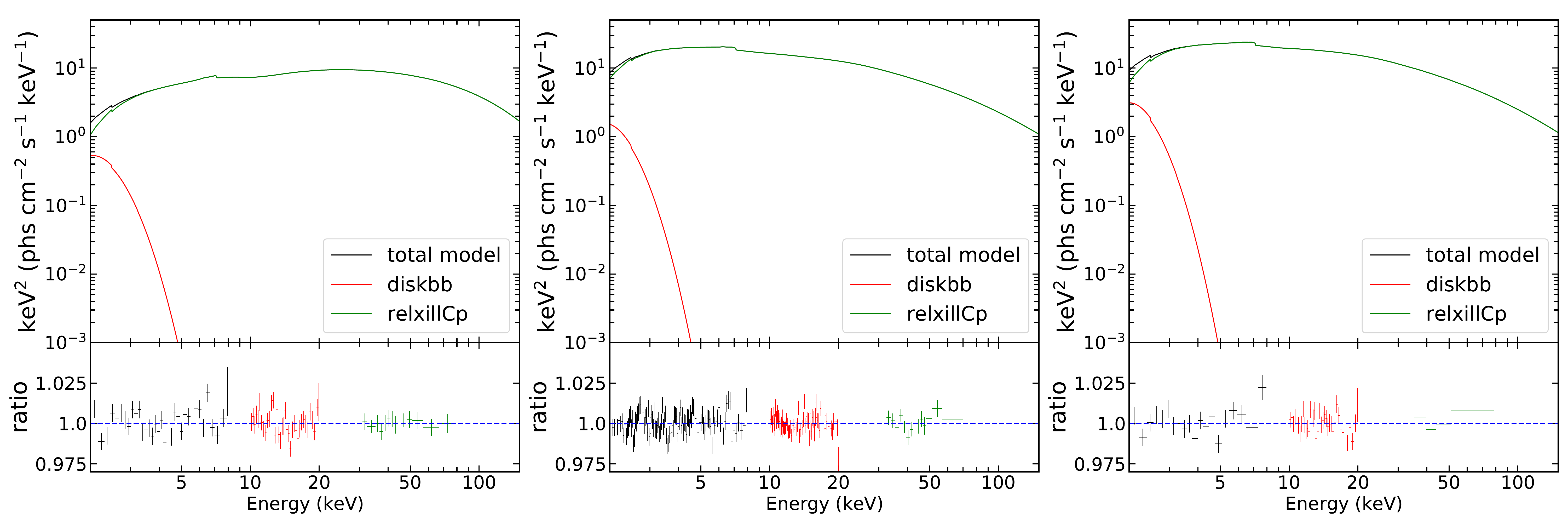}
	\caption{Best-fit models and data to best-fit model ratios for obs~1 (left panel), obs~2 (central panel), and obs~3 (right panel) of Model~2 of MAXI~J1535--571. In the ratio plots, black, red, and green crosses are for LE, ME, and HE data, respectively.}
    \label{eemodel_ratio_MAXI_figure}
\end{figure*}

Tab.~\ref{MAXI_model_2} shows the best-fit values for Model~2. The spectra and the data to best-fit model ratios for obs~1-3 are presented in Fig.~\ref{eemodel_ratio_MAXI_figure}. Compared to Model~1, $\Delta\chi^2=-4036.55$ and $\chi_{\nu}^2=1.06527$. Our measurements of the black hole spin parameter and of the inclination angle of the disk are, respectively, $a_*=0.9916 \pm 0.0012$ and $i=74.2^{+0.6}_{-0.7}$~deg. We note that $N_{\rm H}$ in {\tt tbabs} is left free in the fit and allowed to vary among the three spectra; moreover, we find a higher (about double) value with respect that found in \citet{2018ApJ...860L..28M} with \textsl{NICER} data and in \citet{Sridhar2019} with \textsl{AstroSat} data. 
According to \citet{2009ApJ...707L..77M}, variations in the very soft X-ray band of the spectrum are not due to evolution in neutral photoelectric absorption, even if it is common to allow $N_{\rm H}$ to vary in modeling multiple spectra. However, in MAXI~J1535--571 the inclination angle of the disk is very high, which is not the case for the sources analyzed in \citet{2009ApJ...707L..77M}, so we may have a variable intrinsic absorption due to wind/outflow from the disk. In our fit, {\tt tbabs} is used to model even such a variable intrinsic absorption, which can vary over quite short timescales. In this regard, we note that the estimate of $N_{\rm H}$ in \citet{Xu2018} with \textsl{NuSTAR} data is consistent with our measurements and that the \textsl{NuSTAR} observation analyzed in \citet{Xu2018} overlaps with our obs~1. Following an agnostic approach, we consider also the possibility of a constant $N_{\rm H}$ to check its impact on our measurements.
If we repeat the analysis with $N_{\rm H}$ free but constant over the three observations, we get a slightly worse fit ($\chi^2 = 2704.68$) but the measurements of the model parameters are consistent with those reported in Tab.~\ref{MAXI_model_2}. If we repeat the fit with $N_{\rm H} = 3 \cdot 10^{22}$~cm$^{-2}$ frozen for all observations, we cannot fit the data well ($\chi^2 = 4957.78$) and we have prominent residuals in the LE data.
If the discrepancy between our estimate of $N_{\rm H}$ and that reported by \citet{2018ApJ...860L..28M} and \citet{Sridhar2019} is not related to variable intrinsic absorption, it may be due to some unknown systematic effect, but we note that this does not have a significant impact on our estimates of the black hole spin parameter and inclination angle of the disk.

Last, we consider the possibility of the presence of a distant reflector and we add \texttt{xillverCp} \citep[][]{Garc2010} to the total model of Model~2. We impose that the distant reflector is neutral and we fix $\log{\xi}=0$. However, we find that this new model does not improve the fit and $\chi_{\nu}^2=2666.51/2501=1.06618$, so we do not report here its results and our final model remain Model~2.

\subsection{4U~1630--472}

As in the case of MAXI~J1535--571, even for the spectrum of 4U~1630--472 we start with an absorbed continuum from thermal Comptonization and the XSPEC model is \texttt{constant*tbabs*nthcomp} (Model~0). We find $\chi_{\nu}^2= 9294.76/815=11.4046$.

\begin{figure}
	\includegraphics[width=0.97\columnwidth]{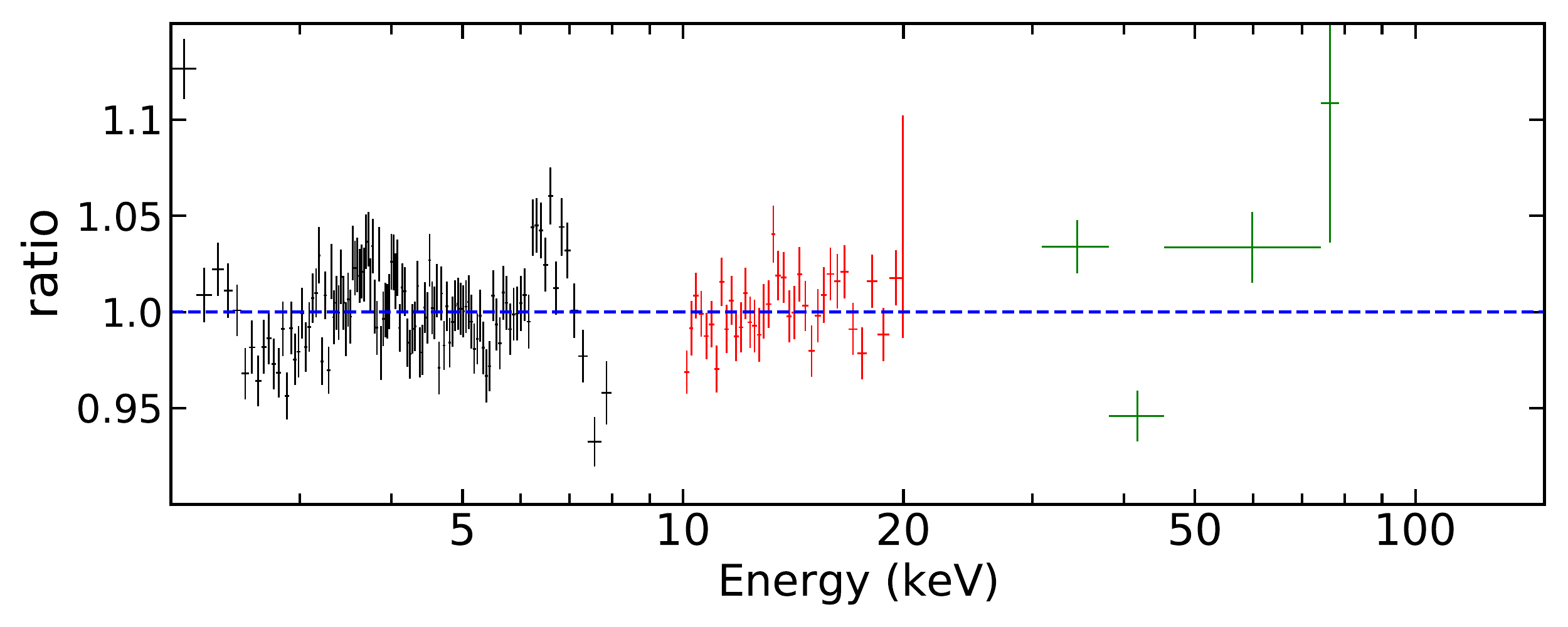}
	\vspace{-0.1cm}
    \caption{Data to best-fit model ratios of Model~1 of 4U~1630--472. Black, red, and green crosses are, respectively, for LE, ME, and HE data.}
    \label{4U_iron_line_figure}
\vspace{0.5cm}
	\includegraphics[width=0.95\columnwidth]{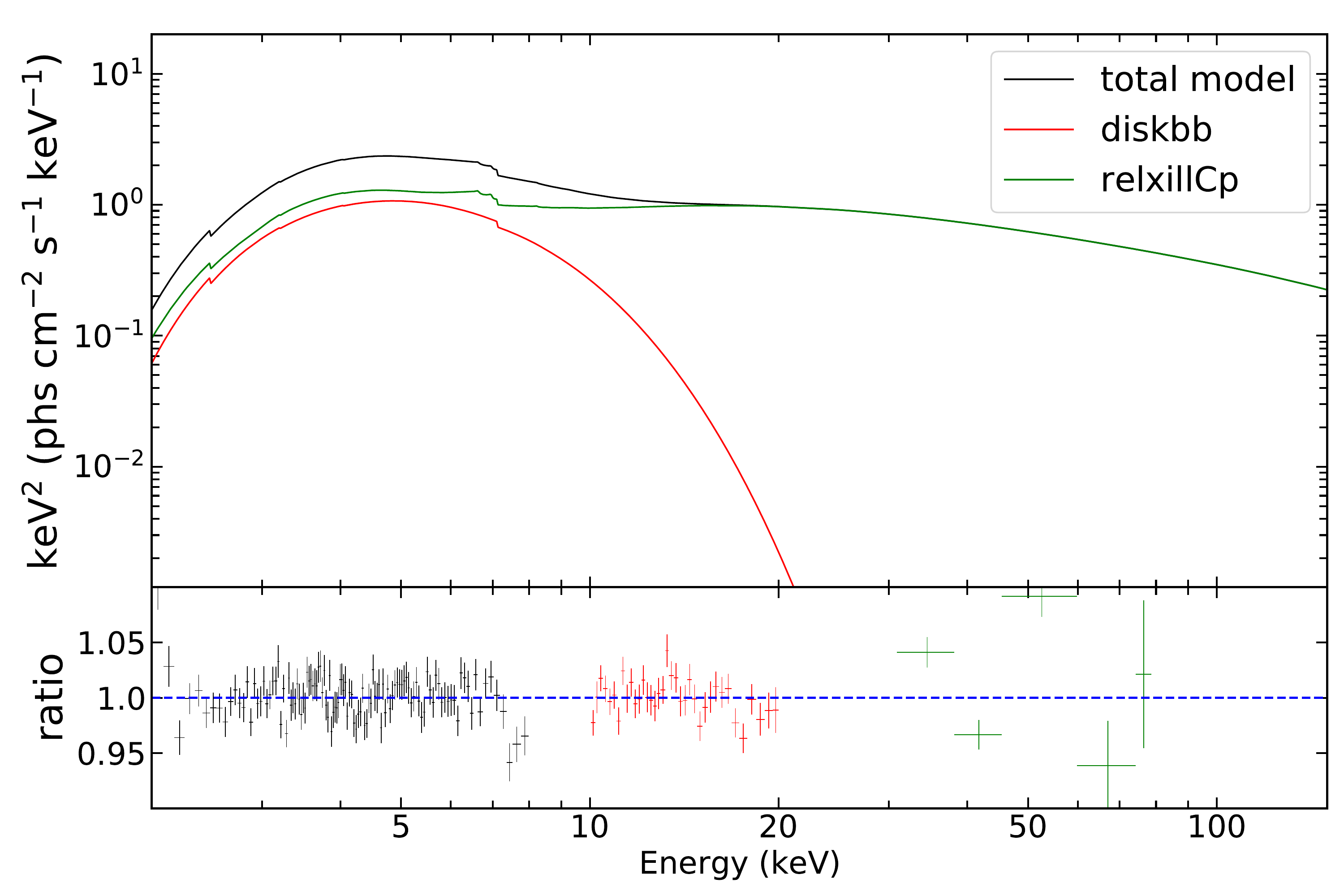}
	\vspace{-0.1cm}
    \caption{Best-fit model and data to best-fit model ratio of Model~2 of 4U~1630--472. In the ratio plots, black, red, and green crosses are for LE, ME, and HE data, respectively.}
    \label{4U_eemod_figure}
\end{figure}

In our Model~1, we add \texttt{diskbb} to include the thermal spectrum from the disk and the fit improves significantly with $\chi_{\nu}^2=1100.88 /814=1.35244$ ($\Delta\chi^2=-8193.88$ with respect to Model~0). The data to best-fit ratio of Model~1 is shown in Fig.~\ref{4U_iron_line_figure}.

For Model~2, we add a relativistically blurred reflection component with $\texttt{relxillCp}$ and our total model becomes $\texttt{constant*tbabs*(diskbb+relxillCp)}$. The emissivity profile of the accretion disk is still modeled with a broken power-law and we have three free parameters in the fit; namely $q_{\rm in}$, $q_{\rm out}$, and $R_{\rm br}$. We find $\chi_{\nu}^2=937.00/804 =1.1654$ ($\Delta\chi^2=-163.88$ with respect to Model~1). Fig.~\ref{4U_eemod_figure} shows the model and the data to best-fit ratio. The estimates of the model parameters are reported in Tab.~\ref{4U_model_2}. Our measurements of the dimensionless black hole spin parameter and inclination angle of the disk are, respectively, $a_*=0.817 \pm 0.014$ and $i= 4^{+6}_{\rm -(B)}$ at 90\% confidence level, where (B) simply indicates that we reach the lower boundary $i = 0$~deg.

\begin{table*}
	\centering
	\renewcommand\arraystretch{1.4}
	\caption{Summary of the best-fit values of Model~2 ($a_*$ and $i$ free; $R_{\rm in}=R_{\rm ISCO}$), Model~2b ($a_*$ free; $i = 64$~deg; $R_{\rm in}=R_{\rm ISCO}$), Model~2c ($a_* = 0.998$; $i$ and $R_{\rm in}$ free), Model~2d (with {\tt relxillD}), and Model~2e (with {\tt xstar}) of 4U~1630--472. 
	The reported uncertainties correspond to the 90\% confidence level for one relevant parameter ($\Delta\chi^2 = 2.71$). $^*$ indicates that the value of the parameter is frozen in the fit. We note that  $A_{\rm Fe}$ is allowed to vary from 0.5 to 10, ${\rm HE}$ {\tt constant} from 0 to 1.2, and $kT_{\rm e}$ from 1 to 400~keV, $\log n_{\rm e}$ from 15 to 19, and $\log\xi$ in {\tt xstar} is from 3 to 6. (B) in the lower uncertainty of the inclination angle means that we reach the lower boundary $i = 0$~deg.}
	\label{4U_model_2}
	\begin{tabular*}{0.80\textwidth}{@{\extracolsep{\fill}}lccccc}
		\hline\hline
		& Model~2 & Model~2b & Model~2c & Model~2d & Model~2e \\
		\hline
		{\tt tbabs} &&&& \\
		$N_{\rm H}$ [10$^{22}$ cm$^{-2}$] & $10.37^{+0.15}_{-0.19}$ & $11.28_{-0.04}^{+0.05}$ & $10.33^{+0.12}_{-0.09}$ & $10.23_{-0.04}^{+0.09}$ & $10.16_{-0.18}^{+0.11}$ \\
		\hline
		{\tt xstar} &&&& \\
		$N_{\rm H}$ [10$^{22}$ cm$^{-2}$] & -- & -- & -- & -- &$6_{-2}^{+9}$\\
		$\log\xi$ [erg~cm~s$^{-1}$] & -- & -- & -- & -- & $4.1_{-0.3}^{+1.9}$\\
		$z$ & -- & -- & -- & -- & $-0.103_{-0.009}^{+0.009}$\\
		\hline
		{\tt diskbb} && \\
		$T_{\rm in}$ [keV] & $1.47^{+0.03}_{-0.03}$ & $1.258_{-0.007}^{+0.004}$ &  $ 1.48 ^{+0.03}_{-0.03}$ & $1.679^{+0.117}_{-0.007}$ & $1.59_{-0.07}^{+0.21}$\\
		${\rm Norm}$ & $51^{+9}_{-5}$ & $146 _{-2}^{+1}$ & $49^{+8}_{-6}$ & $21.1^{+7.2}_{-0.3}$ &$ 33_{-6}^{+8}$\\
		\hline
		{\tt relxillCp} &  &\\
		{\tt relxillD} &  &\\
		$q_{\rm in}$ & $ 6.55^{+0.78}_{-0.14}$ & $10.00_{-0.15}$ & $6.1_{-0.2}^{+0.4} $  & $6.17^{+0.04}_{-0.16}$ & $ 6.7_{-0.3}^{+0.8}$\\
		$q_{\rm out}$ & $2.5^{+0.2}_{-0.2} $ & $0.08_{-0.08}^{+0.12}$ &  $2.48^{+0.26}_{-0.15}$  & $2.84^{+0.04}_{-0.19}$ & $2.6_{-0.3}^{+0.3}$ \\
		$R_{\rm br}$~$[r_{\rm g}]$ & $8.7_{-1.6}^{+1.3}$ & $6.39_{-0.17}^{+0.52}$ &  $8.9_{-1.6}^{+1.8} $ & $8.01^{+0.12}_{-0.86}$ & $8.5_{-1.6}^{+1.6}$\\
		$a_*$ & $0.817_{-0.014}^{+0.014}$ & $0.9932_{-0.0013}^{+0.0016}$ & $0.998^*$ & $0.848_{-0.011}^{+0.004}$ & $0.831_{-0.014}^{+0.014}$\\
		$R_{\rm in}$~$[r_{\rm g}]$ & $R_{\rm ISCO}^*$ & $R_{\rm ISCO}^*$ & $2.58 ^{+0.06}_{-0.07}$ & $R_{\rm ISCO}^*$ & $R_{\rm ISCO}^*$\\
		$i$ [deg] & $4^{+6}_{\rm -(B)}$ & $64^*$ &  $4^{+5}_{\rm -(B)}$ & $3^{+5}_{\rm -(B)}$ & $6_{\rm -(B)}^{+5}$\\
		$A_{\rm Fe}$ & $10.0_{-0.8}$ & $10.0_{-1.2}$ &  $10.0_{-0.4}$ & $10.0_{-0.2}$ & $10.0_{-1.6}$\\
		$\log\xi$ [erg~cm~s$^{-1}$] & $3.94^{+0.08}_{-0.15}$ & $2.98_{-0.17}^{+0.04}$ &  $3.96^{+0.08}_{-0.18}$ & $3.69_{-1.22}^{+0.02}$ & $3.97_{-0.24}^{+0.09}$ \\
		${\Gamma}$ & $2.204^{+0.012}_{-0.011}$ & $2.272_{-0.007}^{+0.005}$ & $2.201_{-0.008}^{+0.063} $ & $2.23_{-0.08}^{+0.03}$ & $2.201_{-0.015}^{+0.034}$\\
		$kT_{\rm e}$ [keV]  & $400_{-198}$ & $132_{-26}^{+55}$ &  $400 _{-186}$ & -- & $400_{-239}$ \\
		$E_{\rm cut}$ [keV]  & -- & -- &  -- & $300^*$ & -- \\
		$\log n_{\rm e}$ [cm$^{-3}$]  & $15^*$ & $15^*$ & $15^*$ & $18.61_{-0.03}^{+0.39}$ & $15^*$  \\
		$R_{\rm f}$ & $0.58^{+0.25}_{-0.09}$ & $0.29_{-0.02}^{+0.02}$ &  $ 0.61 ^{+0.12}_{-0.09}$ & $ 0.85_{-0.03}^{+0.03}$ & $0.71_{-0.13}^{+1.23}$\\
		{Norm} & $0.0141_{-0.0012}^{+0.0009} $  & $0.0197_{-0.0002}^{+0.0002}$ & $0.0137 _{-0.0011}^{+0.0011} $ & $0.013_{-0.002}^{+0.005}$ & $0.0136_{-0.0011}^{+0.0014}$\\
		\hline
		{\tt constant} &&\\
		${\rm LE}$ & $1^*$ & $1^*$ & $1^*$ & $1^*$ & $1^*$ \\
		${\rm ME}$ & $1.10^{+0.03}_{-0.02}$ & $1.166_{-0.009}^{+0.010}$ & $1.102^{+0.03}_{-0.015}$ & $1.095_{-0.018}^{+0.005}$ & $1.05_{-0.03}^{+0.02}$\\
		${\rm HE}$ &$1.2_{-0.02}$ & $1.20_{-0.02}$ & $1.200_{-0.010}$ & $1.200_{-0.009}$ &$1.20_{-0.02}$\\
		\hline
		$\chi^2/\nu$ & 937.00/804 & $1019.71/805$ & 934.52/804 & 926.29/804 & 901.11/801 \\
		 & =1.1654 & =1.2667 & =1.1623 & =1.1521 & =1.1250 \\
		\hline\hline
	\end{tabular*}
\end{table*}

\section{Discussions and conclusions}
\label{ed}

\subsection{MAXI~J1535--571}

In the previous section, we fit the \textsl{Insight-HXMT} spectra of MAXI~J1535--571 assuming a broken power-law emissivity profile. Our final estimates of the black hole spin parameter $a_*$ and of the inclination angle of the disk are (90\% C.L., statistical error)
\be
a_* = 0.9916 \pm 0.0012 \, , \quad
i = 74.2^{+0.6}_{-0.7}~{\rm deg} \, . \nonumber
\ee
These measurements can be compared with those reported in \citet{Xu2018} from the analysis of a \textsl{NuSTAR} observation on 7~September~2017, so overlapping our obs~1. For a broken power-law emissivity profile with $q_{\rm out}$ fixed to 3 and $R_{\rm br}$ fixed to 10~$r_{\rm g}$, \citet{Xu2018} obtain 
\be
a_* > 0.987 \, , \quad i = 75^{+2}_{-4}~{\rm deg} \, , \nonumber
\ee
which are in perfect agreement with our \textsl{Insight-HXMT} estimates. On the other hand, there are some discrepancies in the estimates of the other model parameters between our obs~1 and \citet{Xu2018}. In \citet{Xu2018}, $kT_{\rm in} = 0.40 \pm 0.01$~keV in {\tt diskbb} and $kT_{\rm e} = 21.9 \pm 1.2$~keV and $R_{\rm f} = 0.60_{-0.10}^{+0.06}$ in {\tt relxillCp}, while the photon index $\Gamma$ and the ionization parameter $\xi$ are consistent with our estimates. However, there are also some differences in the model. First, \citet{Xu2018} assume $q_{\rm out} = 3$ and $R_{\rm br} = 10~r_{\rm g}$, so the only free parameter of the emissivity profile is $q_{\rm in}$. Second, their model includes a distant reflector described by {\tt xillverCp}. The inner edge of the disk is free in \citet{Xu2018}, not like in our case fixed to the ISCO radius, but they find it is consistent with the ISCO radius.

\citet{Xu2018} consider even a lamppost emissivity profile \citep{Dauser2013} and obtain a slightly higher $\chi^2$ ($\Delta\chi^2 = 23$) with one less free parameter. For the lamppost model, their measurements of the black hole spin parameter and of the inclination angle of the disk are, respectively, $a_*>0.84$ and $i = 57_{-2}^{+1}$~deg. Fitting our \textsl{Insight-HXMT} data with the lamppost model for the three spectra (not only obs~1), we find a worse fit with $\chi_{\nu}^2=2864.76/2508=1.14225$. The measurement of the black hole spin parameter and of the inclination angle of the disk are now $a_* > 0.91$ and $i = 37.0_{-1.4}^{+2.1}$.

We note that there are other three spin measurements of MAXI~J1535--571 reported in the literature. From the analysis of \textsl{NICER} data, \citet{2018ApJ...860L..28M} find $a_* = 0.994 \pm 0.002$ (1-$\sigma$), which is consistent with our measurement and that in \citet{Xu2018}. From the analysis of \textsl{AstroSat} data, \citet{Sridhar2019} find $a_* = 0.67_{-0.04}^{+0.16}$ employing the lamppost emissivity profile of {\tt relxillCp}. They also find an inclination angle around 40~deg when they do not include the distant reflector and around 80~deg with the distant reflector. From the spectral analysis of the \textsl{Insight-HXMT} observation P011453500107 (one of the five observations in our obs~1 of MAXI~J1535--571, see Tab.~\ref{observation_table}), \citet{2020JHEAp..25...29K} infer the black hole spin parameter $a_* = 0.7 _{-0.3}^{+0.2}$ (90\% C.L.) employing the lamppost version of {\tt relxillCp}. However, \citet{2020JHEAp..25...29K} is mainly devoted to the timing analysis of the \textsl{Insight-HXMT} observations of MAXI~J1535--571 and the measurement of the black hole spin is inferred from a short observation with a low number of photon counts.

For obs~2 and obs~3, we find that $q_{\rm in}$ is stuck at the upper boundary of the parameter range and the value of $q_{\rm out}$ is quite low. Such an emissivity profile is not rare in Galactic black holes; see, for instance, the case of GS~1354--645~\citep{2016ApJ...826L..12E,2018ApJ...865..134X} and of GRS~1915+105~\citep{2013ApJ...775L..45M,2019ApJ...884..147Z}. Even \citet{2018ApJ...860L..28M} found a very high $q_{\rm in}$ and a very low $q_{\rm out}$ in MAXI~J1535--571 with \textsl{NICER} data. These emissivity profiles may be generated by coronae with ring-like geometries located above the accretion disk \citep[][]{2003MNRAS.344L..22M,2015MNRAS.449..129W,2020arXiv201207469R}. For similar coronae, we could indeed expect a very steep emissivity profile near the inner edge of the accretion disk, an almost-flat emissivity profile in an intermediate region of the accretion disk, and an emissivity index $q$ close to 3 at larger radii. However, since the flux from large radii is low, we may not be able to recover the emissivity index $q = 3$ of the outer part of the accretion disk from the fit\footnote{In our fit, the outer radius of the disk is fixed at $R_{\rm out} = 400$~$r_{\rm g}$. We find that the fluxes from the inner region $R_{\rm in} < r < R_{\rm br}$ and from the outer region $R_{\rm br} < r < R_{\rm out}$ are comparable in both obs~2 and obs~3. If we set $R_{\rm out} = 1000$~$r_{\rm g}$, the result of our fit does not change ($\Delta\chi^2 = -19$ with respect to Model~2) and the fluxes from the inner and outer regions are still comparable, so the choice of the value of $R_{\rm out}$ does not affect our measurement. If we employ a broken power law for obs~1 and a twice broken power law for obs~2 and obs~3, always imposing that the outer emissivity index is 3, we do not improve the fit ($\Delta\chi^2 = -20$ with respect to Model~2); the breaking radii of the outer part of the disk are a few hundreds gravitational radii, but their values cannot be constrained well.}. We also note that in obs~2 and obs~3 the photon index $\Gamma$ is around 2.6 (it is $\sim 1.9$ in obs~1) and the coronal temperature is significantly higher than in obs~1. The differences in the emissivity profiles and in the values of the photon index and of the electron temperature between obs~1 and obs~2/obs~3 suggest important changes in the properties of the corona between 7~September~2017 and 12~September~2017.

We note that we find a higher value of the ionization parameter $\xi$ in obs~1 and a lower value in obs~2 and obs~3, while the flux of the reflection component is higher in obs~2 and obs~3 and lower in obs~1. The ionization parameter is defined as
\be
\xi = \frac{4 \pi F_X}{n_{\rm e}} \, ,
\ee
where $F_X$ and $n_{\rm e}$ are, respectively, the X-ray flux illuminating the disk and the electron density of the disk. Since the intensity of the reflection component is approximately proportional to $F_X$, we should expect that $\xi$ increases from obs~1 to obs~2/obs~3, while it is exactly the opposite and $\xi$ decreases. 
As we have discussed in the previous paragraph, the coronal geometry must change from obs~1 to obs~2/obs~3, and this has certainly some impact on the value of $\xi$. The intensity of the thermal spectrum (norm in {\tt diskbb} in our fit) increases by more than an order of magnitude from obs~1 to obs~2/obs~3, but this should actually lead to reduce the value of the disk electron density $n_{\rm e}$ \citep[see, e.g.,][]{2019MNRAS.484.1972J} and further increase the value of the ionization $\xi$.

To explore further the trend of the ionization parameter, we replace {\tt relxillCp} with {\tt relxillD} (Model~2d), which permits us to have a free disk electron density $n_{\rm e}$ in the fit (but, unfortunately, the coronal spectrum is described by a power law with fixed high-energy cutoff $E_{\rm cut} = 300$~keV). The results of our fit are shown in Tab.~\ref{MAXI_model_2}. With Model~2d, we recover the expected trend in the ionization parameter, whose value increases from obs~1 to obs~3. 
The disk electron density $n_{\rm e}$ increases from obs~1 to obs~2/obs~3, which is not the prediction for a radiation-pressure dominated standard accretion disk model~\citep[][]{1994ApJ...436..599S,2019MNRAS.484.1972J}.
We note, however, that the estimates of the parameters in obs~1 should be taken with caution because with Model~2 we find a low coronal temperature for obs~1 and therefore the high-energy cutoff $E_{\rm cut} = 300$~keV in {\tt relxillD} certainly introduces some undesirable bias in the estimate of some model parameters.

For the iron abundance $A_{\rm Fe}$, we find quite a low value. The reliability of the iron abundance measurements inferred from the analysis of the reflection spectrum of accretion disks is a well-known issue~\citep[see, e.g.,][and references therein]{2021SSRv..217...65B}. For most sources, we recover unreasonably high value of $A_{\rm Fe}$, as we find in our analysis of 4U~1630--472 (its discussion is in the next subsection). However, for some sources we have the opposite problem and the measurement of the iron abundance is unexpectedly low \citep[see, e.g.,][]{2016ApJ...826L..12E,2018ApJ...865...18X,2019ApJ...884..147Z,2020ApJ...900...78D}. Low iron abundances are found when the black hole spin parameter and the inclination angle of the disk are very high, which is probably the case even of MAXI~J1535--571. According to the study reported in \citet{2021ApJ...910...49R}, the iron abundance would be indeed underestimated in similar sources when the reflection model of the fit does not include the returning radiation, namely the radiation emitted by the disk and returning to the the disk because of the strong light bending near the black hole. While such an explanation could fit well even with MAXI~J1535--571, the conclusions reported in \citet{2021ApJ...910...49R} should be taken with some caution, as their simulations are based on a model with a neutral disk, while real accretion disks are highly ionized.

As the thermal photons from the inner part of the accretion disk can inverse Compton scatter off free electrons in the corona, even the reflection spectrum can be affected by Compton scattering in the corona. To check if the Compton scattering of the reflection photons has an impact on our results, we fit the data with the model {\tt constant*tbabs*(diskbb + nthcomp + simplcutx*relxillCp)}, where {\tt simplcutx} takes into account the Comptonization of the reflection photons~\citep{2017ApJ...836..119S} and we set to $-1$ the reflection fraction $R_{\rm f}$ in {\tt relxillCp} (the spectrum of the corona is now describes by {\tt nthcomp}). Such a model provides a slightly better fit, $\chi^2 = 2658.53$ ($\Delta\chi^2 = - 8.90$ with respect to Model~2), but the scattering fraction in {\tt simplcutx} is low in all observations ($f_{\rm SC} = 0.15-0.25$) and we do not see any significant difference in the estimate of the model parameters with respect to the measurements of Model~2.

\begin{figure}
	\includegraphics[width=0.95\columnwidth]{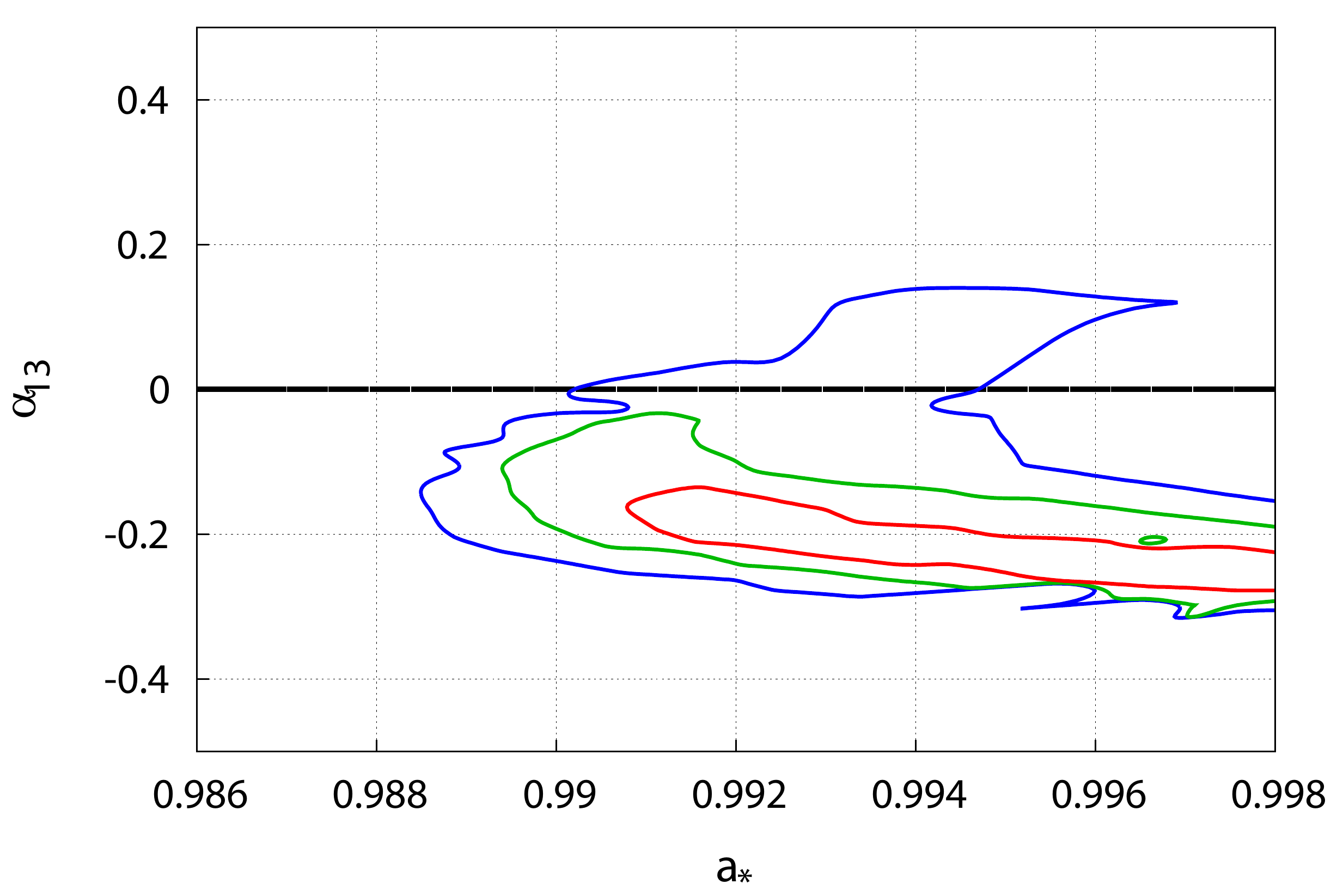}
	\vspace{-0.1cm}
    \caption{Constraints on the spin parameter $a_*$ and on the Johannsen deformation parameter $\alpha_{13}$ from Model~3 of MAXI~J1535--571. The red, green, and blue lines indicate, respectively, the 68\%, 90\%, and 99\% confidence level contours for two relevant parameters ($\Delta\chi^2 = 2.30$, 4.61, and 9.21, respectively). }
    \label{maxij_contour}
\end{figure}

Last, considering the good quality of these \textsl{Insight-HXMT} data, we replace \texttt{relxillCp} with \texttt{relxillCp\_nk} \citep{2017ApJ...842...76B,Abdikamalov2019} (Model~3), which is designed to test the Kerr spacetime around astrophysical black holes~\citep[see, e.g.,][]{Tripathi2021a,2021arXiv210604084B}. In \texttt{relxillCp\_nk}, the spacetime metric includes some ``deformation parameters'' that are introduced to quantify possible deviations from the Kerr solution. The Kerr metric is recovered when all deformation parameters vanish and, as in a null experiment, the spirit is to check whether observations require vanishing deformation parameters. Here we employ the version of the model in which deviations from the Kerr background are described by the deformation parameter $\alpha_{13}$ of the Johannsen metric \citep{2013PhRvD..88d4002J}. We refit the data with $\alpha_{13}$ free ($\alpha_{13} = 0$ for the Kerr metric). The best-fit has $\chi^2_{\nu}= 2663.40/2503=1.06408$ and improves only marginally the fit of Model~2 ($\Delta\chi^2=-4.03$). The estimates of the other model parameter do not show any significant change. The best-fit values of Model~3 are reported in the right part of Tab.~\ref{MAXI_model_2}. The constraints on the black hole spin parameter $a_*$ and on the deformation parameter $\alpha_{13}$ after marginalizing over all other free parameters are shown in Fig.~\ref{maxij_contour}. The Kerr metric is recovered at a confidence level a bit larger than 2-$\sigma$, where $\Delta\chi^2=4$. The 3-$\sigma$ measurement of the Johannsen deformation parameter $\alpha_{13}$ is (statistical error only) 
\be\label{eq-a13-gr}
\alpha_{13} = -0.24_{-0.07}^{+0.36} \, .
\ee
The most stringent constraint on $\alpha_{13}$ to date has been obtained from the analysis of a \textsl{NuSTAR} observation of the black hole in GX~339--4 in \citet{2021ApJ...907...31T} and the 3-$\sigma$ measurement is
\be
\alpha_{13} = -0.02_{-0.14}^{+0.03} \, .
\ee
Our \textsl{Insight-HXMT} constraint on $\alpha_{13}$ in Eq.~(\ref{eq-a13-gr}) is not too far from the most stringent \textsl{NuSTAR} constraint, presumably thanks to the better energy resolution of \textsl{Insight-HXMT} near the iron line, despite a significantly lower effective area with respect to the \textsl{NuSTAR} detectors.

\subsection{4U~1630--472}

For 4U~1630--472, we find the following estimates of the black hole spin parameter $a_*$ and inclination angle of the disk $i$ (90\% C.L., statistical error)
\be
a_*=0.817 \pm 0.014 \, , \quad i=4^{+6}_{\rm -(B)}~{\rm deg} \, .
\ee
From the \textsl{NuSTAR} observation on 9~March~2013, so in some previous outburst of the source, \citet{King2014} find the following estimates of $a_*$ and $i$
\be
a_*=0.985_{-0.014}^{+0.005} \, , \quad i=64^{+2}_{-3}~{\rm deg} \, ,
\ee
where here the uncertainties are at 1-$\sigma$. \citet{King2014} employ the reflection model {\tt refbhb}~\citep{2007MNRAS.381.1697R,2008MNRAS.387.1489R}, which is designed for modeling the reflection spectrum of Galactic black holes in the soft or intermediate states, when the thermal soft X-ray photons of the accretion disk can contribute to the reflection spectrum. With the same \textsl{NuSTAR} data and employing the lamppost version of {\tt relxillCp}, \citet{Tripathi2021a} still find a very high black hole spin parameter, $a_*=0.990_{-0.005}^{+0.003}$, even if the inclination angle of the disk is $i=8^{+6}_{-2}$ and consistent with our \textsl{Insight-HXMT} measurement. So the reflection model does not seem to be responsible for the discrepancy in the black hole spin parameter while it may have a role in the estimate of the inclination angle of the disk. Since the \textsl{NuSTAR} and \textsl{Insight-HXMT} observations are not simultaneous, as it was in the case of MAXI~J1535--571 for obs~1, we cannot compare the estimates of the other model parameters, which are expected to change rapidly. 

\begin{figure}
	\includegraphics[width=0.97\columnwidth]{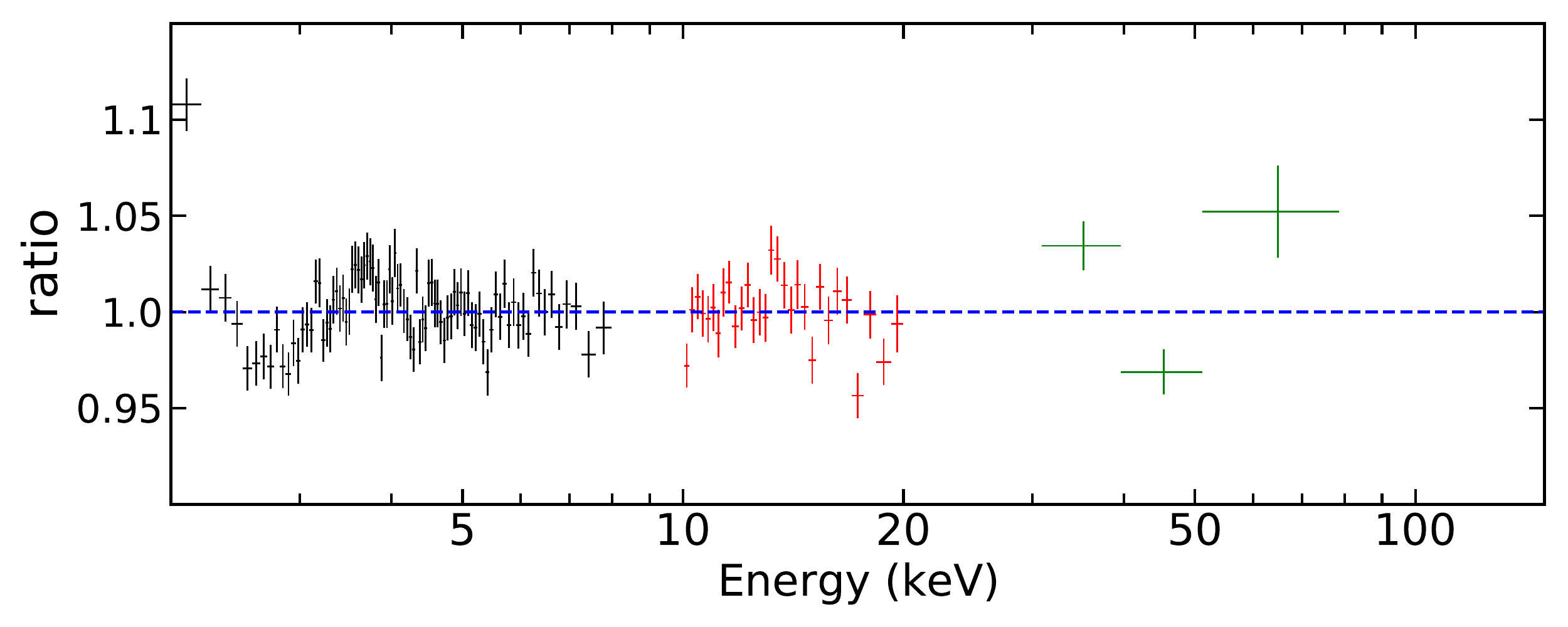}
	\vspace{-0.1cm}
    \caption{Data to best-fit model ratios of Model~2b of 4U~1630--472. Black, red, and green crosses are, respectively, for LE, ME, and HE data.}
    \label{4U_inclination_64_figure}
\end{figure}

A high value of the inclination angle of the disk is suggested by the presence of X-ray dips \citep{1998ApJ...494..753K} and by the shape of the HID \citep{2013MNRAS.432.1330M}, which would be consistent with the result in \citet{King2014}. We thus refit the data with the inclination angle frozen to the best-fit of \citet{King2014}. This is our Model~2b and the results are shown in the third column of Tab.~\ref{4U_model_2}. The fit is certainly worse ($\Delta\chi^2 = 82.71$ with respect to Model~2). Fig.~\ref{4U_inclination_64_figure} shows the ratio plot of Model~2b and we clearly see residuals in the LE data. However, we find a spin parameter close to 1, $a_* = 0.9932_{-0.0013}^{+0.0016}$, and consistent with the estimate of \citet{King2014}.

From the analysis of \textsl{AstroSat} and \textsl{Chandra} data, \citet{2018ApJ...867...86P} report the black hole spin parameter $a_* = 0.92 \pm 0.04$ (99.7\% C.L.). Such a measurement is obtained from the analysis of the thermal spectrum of the disk, not from that of the reflection features of the disk as in our work and in \citet{King2014}. Spin measurements of 4U~1630--472 from the analysis of reflection features in \textsl{NuSTAR} and \textsl{Swift}/XRT data are reported even in \citet{2021ApJ...909..146C}. However, the authors employ different models and eventually get two possible values of the black hole spin: $a_* = 0.989_{-0.002}^{+0.001}$ or $a_* = 0.85 \pm 0.07$. The latter measurement, which is statistically preferred, would be consistent with our measurement from \textsl{Insight-HXMT} data.

A lower value of the black hole spin parameter can be easily obtained in the case of a truncated disk, namely if the inner edge of the disk is not at the ISCO but at some larger radius. To check whether the accretion disk could be truncated during the \textsl{Insight-HXMT} observation on 19-20~March~2020, we refit the data assuming $a_* = 0.998$ and leaving the inner edge of the disk $R_{\rm in}$ free in the fit (Model~2c). The results of the new fit are reported in Tab.~\ref{4U_model_2}. The fit is slightly better ($\Delta\chi^2 = -2.48$) and we find $R_{\rm in} = 2.58_{-0.07}^{+0.06}~r_{\rm g}$. However, assuming that the black hole mass is 10~$M_\odot$~\citep{2014ApJ...789...57S} and that the distance of the source from us is 10~kpc~\citep{2001A&A...375..447A}, the Eddington-scaled disk luminosity during the \textsl{Insight-HXMT} observation is very close to 30\%. From the HID, we also see that the source was in an intermediate state. This is not the situation in which we would expect a truncated disk, so it is not natural to explain the discrepancy between our spin measurement with that in \citet{King2014} with a truncated disk in the \textsl{Insight-HXMT} data.

Our estimate of the iron abundance $A_{\rm Fe}$ is very high. This is a common issue in the analysis of the reflection spectrum of many sources and there is no common consensus on its origin. It is often thought that such super-solar iron abundances are the result of some deficiency in current reflection models \citep[see, e.g., the list of possible explanations in][]{2021SSRv..217...65B}. However, even real physical mechanisms cannot be excluded, and an example is the radiative levitation of metal ions in the inner part of the accretion disk~\citep{2012ApJ...755...88R}.

As in the case of MAXI~J1535--571, we explore the impact of a higher disk electron density and we replace {\tt relxillCp} with {\tt relxillD} (Model~2d). The results of our fit are shown in the fifth column of Tab.~\ref{4U_model_2}. The fit is slightly better ($\Delta\chi^2 = -10.71$ with respect to Model~2) and we recover a significantly higher disk electron density than the value in {\tt relxillCp}, $\log n_{\rm e} = 18.61_{-0.03}^{+0.39}$. A higher disk electron density has some weak impact on the estimate of the parameters in {\tt diskbb}, where we find a higher temperature and a lower normalization, while we do not see significant changes in the estimate of the other parameters. In particular, the estimate of the black hole spin parameter and of the inclination angle of the disk are consistent with those of Model~2.

Last, we check if we can describe better the absorption of the source by adding \texttt{xstar} (Model~2e). We assume solar metal abundances. The results are reported in the last column of Tab.~\ref{4U_model_2}. The fit is a bit better with $\chi^2_{\nu}=900.70/801=1.1245$ ($\Delta\chi^2 = - 36.3$ with respect to Model~2), but the measurements of the spin parameter and of the inclination angle do not change significantly: $a_*=0.833_{-0.013}^{+0.005}$ and $i=3^{+7}_{\rm -(B)}$. Even the estimates of the other parameters in {\tt relxillCp} and {\tt diskbb} are consistent with those of Model~2. Concerning the parameters in {\tt xstar}, we find the column density $N_{\rm H}=6^{+9}_{-2} \cdot 10^{22}$~cm$^{-2}$, the ionization $\log\xi=4.1^{+0.2}_{-0.3}$, and the outflow blueshift $z=-0.103\pm 0.009$.
We note that a similar blueshift corresponds to an outflow velocity of 10\% of the speed of light, which is certainly unusual for a black hole binary \citep[but see][]{2018ApJ...865...18X}. However, it is difficult to simultaneously constrain the ionization $\xi$ and the outflow velocity, because these two parameters are degenerate in the \texttt{xstar} grid. Without an estimate of the ionized outflow mass, we cannot check whether our result is energetically allowed. On the other hand, similar -- and often higher -- outflow velocities are commonly seen in ultra-luminous X-ray sources~\citep{2017AN....338..234P} and active galactic nuclei~\citep{2010A&A...521A..57T,2013MNRAS.430...60G}, so we cannot exclude a similar phenomenon in our \textsl{Insight-HXMT} observation of 4U~1630--472. It is not the purpose of the present work to investigate the properties of such an outflow, and here we want only to stress that including \texttt{xstar} in the model does not change our estimates of the black hole spin and of the inclination angle of the disk.


\section*{Acknowledgements}

This work was supported by the National Natural Science Foundation of China (NSFC), Grant No.~11973019, the Natural Science Foundation of Shanghai, Grant No. 22ZR1403400, the Shanghai Municipal Education Commission, Grant No. 2019-01-07-00-07-E00035, Fudan University, Grant No.~JIH1512604, and Fudan's Undergraduate Research Opportunities Program (FDUROP). 
The work of J.L. was supported by the National Natural Science Foundation of China (NSFC), Grant No.~12173103, 11733009, U2038101, and U1938103, and the Guangdong Major Project of the Basic and Applied Basic Research, Grant No.~2019B030302001.


\section*{Data Availability}

The \textsl{Insight-HXMT} raw data analyzed in this work are available to download at the IHEP website \href{http://hxmten.ihep.ac.cn}{http://hxmten.ihep.ac.cn}.


\bibliographystyle{mnras}
\bibliography{ref} 


\bsp	
\label{lastpage}

\end{document}